\documentclass[apjl]{emulateapj}

\usepackage{graphicx}
\usepackage{amssymb}
\usepackage{txfonts}
\usepackage{natbib}

\shorttitle{Discovery of very high energy $\gamma$-ray emission from Centaurus A}
\shortauthors{Aharonian et al. (H.E.S.S. Collaboration)}

\begin{document}

\title{Discovery of very high energy $\gamma$-ray emission from Centaurus A with H.E.S.S.}

%

\author{F. Aharonian~\altaffilmark{1,13},
  A.G.~Akhperjanian~\altaffilmark{2},
  G.~Anton~\altaffilmark{16},
  U.~Barres de Almeida~\altaffilmark{8,a},
  A.R.~Bazer-Bachi~\altaffilmark{3},
  Y.~Becherini~\altaffilmark{12},
  B.~Behera~\altaffilmark{14},
  W.~Benbow~\altaffilmark{31},
  K.~Bernl\"ohr~\altaffilmark{1,5},
  C.~Boisson~\altaffilmark{6}
% ...
}

% ...
\altaffiltext{1}{
Max-Planck-Institut f\"ur Kernphysik, P.O. Box 103980, D 69029
Heidelberg, Germany}
\altaffiltext{2}{
 Yerevan Physics Institute, 2 Alikhanian Brothers St., 375036 Yerevan,
Armenia}
\altaffiltext{3}{
Centre d'Etude Spatiale des Rayonnements, CNRS/UPS, 9 av. du Colonel Roche, BP
4346, F-31029 Toulouse Cedex 4, France}
% ...
\altaffiltext{5}{
Institut f\"ur Physik, Humboldt-Universit\"at zu Berlin, Newtonstr. 15,
D 12489 Berlin, Germany}
\altaffiltext{6}{
LUTH, Observatoire de Paris, CNRS, Universit\'e Paris Diderot, 5 Place Jules Janssen, 92190 Meudon, 
France}
% ...
\altaffiltext{8}{
University of Durham, Department of Physics, South Road, Durham DH1 3LE,
U.K.}
% ...
\altaffiltext{12}{
Astroparticule et Cosmologie (APC), CNRS, Universite Paris 7 Denis Diderot,
10, rue Alice Domon et Leonie Duquet, F-75205 Paris Cedex 13, France}
\altaffiltext{13}{
Dublin Institute for Advanced Studies, 5 Merrion Square, Dublin 2,
Ireland}
\altaffiltext{14}{
Landessternwarte, Universit\"at Heidelberg, K\"onigstuhl, D 69117 Heidelberg, Germany}
% ...
\altaffiltext{16}{
Universit\"at Erlangen-N\"urnberg, Physikalisches Institut, Erwin-Rommel-Str. 1,
D 91058 Erlangen, Germany}
% ...
\altaffiltext{31}{Fred Lawrence Whipple Observatory, Harvard-Smithsonian Center for Astrophysics, Amado, AZ85645, USA}
\altaffiltext{a}{supported by CAPES Foundation, Ministry of Education of Brazil}


\begin{abstract}
We report the discovery of faint very high energy (VHE, $E > 100$\,GeV) $\gamma$-ray emission from the radio galaxy Centaurus A in observations performed with the H.E.S.S. experiment, an imaging atmospheric Cherenkov telescope array consisting of four telescopes located in Namibia. Centaurus A has been observed for more than 120\,h.
A signal with a statistical significance of 5.0\,$\sigma$ is detected from the region including the radio core and the inner kpc jets.
The integral flux above an energy threshold of $\sim$250\,GeV is measured to be $~0.8\,\%$ of the flux of the Crab Nebula (apparent luminosity: L($>$250\,GeV)$\approx2.6 \times 10^{39}$\,erg\,s$^{-1}$, adopting a distance of 3.8\,Mpc). The spectrum can be described by a power law with a photon index of $2.7 \pm 0.5_{\mathrm{stat}} \pm 0.2_{\mathrm{sys}}$. No significant flux variability is detected in the data set. However, the low flux only allows detection of variability on the timescale of days to flux increments above a factor of $\sim15-20$ ($3\,\sigma$ and $4\,\sigma$, respectively). The discovery of VHE $\gamma$-ray emission from Centaurus A reveals particle acceleration in the source to $>$\,TeV energies and, together with M\,87, establishes radio galaxies as a class of VHE emitters.
\end{abstract}

\keywords{galaxies: active --- galaxies: individual (\objectname{Cen A}) --- gamma rays: observations}


\section{Introduction}

Centaurus~A (Cen~A) is the nearest active galaxy (for a review see \citealt{israel:1998a}). It is classified as a FR I radio galaxy; these are thought to be the parent population of BL Lac objects \citep{urry:1995a}. At radio wavelengths rich jet structures are visible, extending from the core and the inner pc and kpc jet to giant outer lobes with an angular extension of $8^{\circ} \times 4^{\circ}$. The inner kpc jet has also been detected in X-rays, revealing a complex structure of bright knots and diffuse emission \citep{kraft:2002a}. The angle of the jet axis to the line of sight is estimated to be 15-80$^\circ$ \citep[see e.g.][and references therein]{horiuchi:2006a}. With a distance of 3.8\,Mpc (\citealt{rejkuba:2004a}; $1'$ corresponds to $\sim1.1$\,kpc projected length) even the inner jet structures are resolvable with the angular resolution of current experiments in the very high energy (VHE; $E > 100$\,GeV) regime\footnote{H.E.S.S. angular resolution: $\sim6'$ per event,  $\sim6-30''$ systematic error on position depending on the dataset \citep{vaneldik:2007a:hess:icrc:sgra}.}. The elliptical host NGC~5128 features a dark lane, a thin edge-on disk of dust and young stars, believed to be the remnant of a merger. Recent estimates for the mass of the central supermassive black hole give $ (5.5 \pm 3.0)  \times 10^{7}$\,$M_\odot$ \citep{cappellari:2008a}.
The kpc-scale jet and the active nucleus are confirmed sources of strong non-thermal emission. In addition, more than 200 X-ray point sources with an integrated luminosity of L$_{\mathrm{X}} > 10^{38}$\,erg\,s$^{-1}$ are established to be associated with the host galaxy \citep{kraft:2001a}.
Recently, \citet{croston:2009a} reported the detection of non-thermal X-ray emission from the shock of the southwest inner radio lobe from deep Chandra observations.

Cen~A was detected at MeV to GeV energies by all instruments on board the Compton Gamma-Ray Observatory (CGRO) in the period 1991 -- 1995, revealing a peak in the spectral energy distribution (SED) in $\nu F_\nu$ representation at  $\sim0.1$\,MeV with a maximum flux of about $\sim10^{-9}$\,erg\,cm$^{-2}$\,s$^{-1}$  \citep{steinle:1998a}.  \citet{steinle:1998a} also reported variability of the Cen~A source, especially pronounced at 10\,MeV, while \citet{sreekumar:1999a} found the EGRET flux was stable during the whole period of CGRO observations. A tentative detection of Cen~A ($4.5\sigma$) at VHE during a giant X-ray outburst in the 1970's was reported  by \citet{grindlay:1975a}. Subsequent VHE observations made with the Mark~III \citep{carraminana:1990a}, JANZOS \citep{allen:1993a}, CANGAROO \citep{rowell:1999a, kabuki:2008a}, and H.E.S.S. \citep{aharonian:2005:hess:agnul} experiments resulted in upper limits.

Cen~A has been proposed as a possible source of ultra-high energy cosmic rays (UHECR; E $> 6 \times 10^{19-20}$\,eV) (\citealt{romero:1996a}, but see also \citealt{lemoine:2008a}). Recently, the Pierre Auger Collaboration reported an anisotropy in the arrival direction of UHECR \citep{augercollaboration:2007a,augercollaboration:2008a}. While a possible correlation with nearby active galactic nuclei is still under discussion, it has been pointed out that several of the events can possibly be associated with Cen~A \cite[e.g.][]{gorbunov:2008a, moskalenko:2008a,kachelriess:2008a}.

Until now, the only firmly established extragalactic VHE $\gamma$-ray source with only weakly beamed emission is the giant radio galaxy M\,87 \citep{aharonian:2003b,aharonian:2006:hess:m87:science}. M\,87 showed strong flux outbursts in the VHE regime with timescales on the order of days \citep{aharonian:2006:hess:m87:science,albert:2008:magic:m87}, pointing to a characteristic size of the emission region $< 5 \times 10^{15} \delta$\,cm, corresponding to $\approx 5\delta$ Schwarzschild radii ($M_{\mathrm{BH}} = 3.2 \times 10^{9}$\,M$_{\odot}$, $\delta$: relativistic Doppler factor).
Recently, \citet{aliu:2009a:magic:3c66ab} and \citet{acciari:2009a:veritas:3c66a} reported VHE emission from the direction of the blazar 3C\,66\,A and the radio galaxy 3C\,66\,B (angular separation 6'). While  \citet{aliu:2009a:magic:3c66ab} favor 3C\,66\,B as the origin of the VHE emission in their dataset, \citet{acciari:2009a:veritas:3c66a} exclude 3C\,66\,B as the origin of their detected emission with a significance of 4.3$\sigma$.

In this Letter the discovery of VHE emission from Cen~A with the H.E.S.S. experiment is reported, and properties of the detected emission and their implications are discussed.


\section{H.E.S.S. observations and results}

\begin{figure}
\centering
\includegraphics[width=0.5\textwidth]{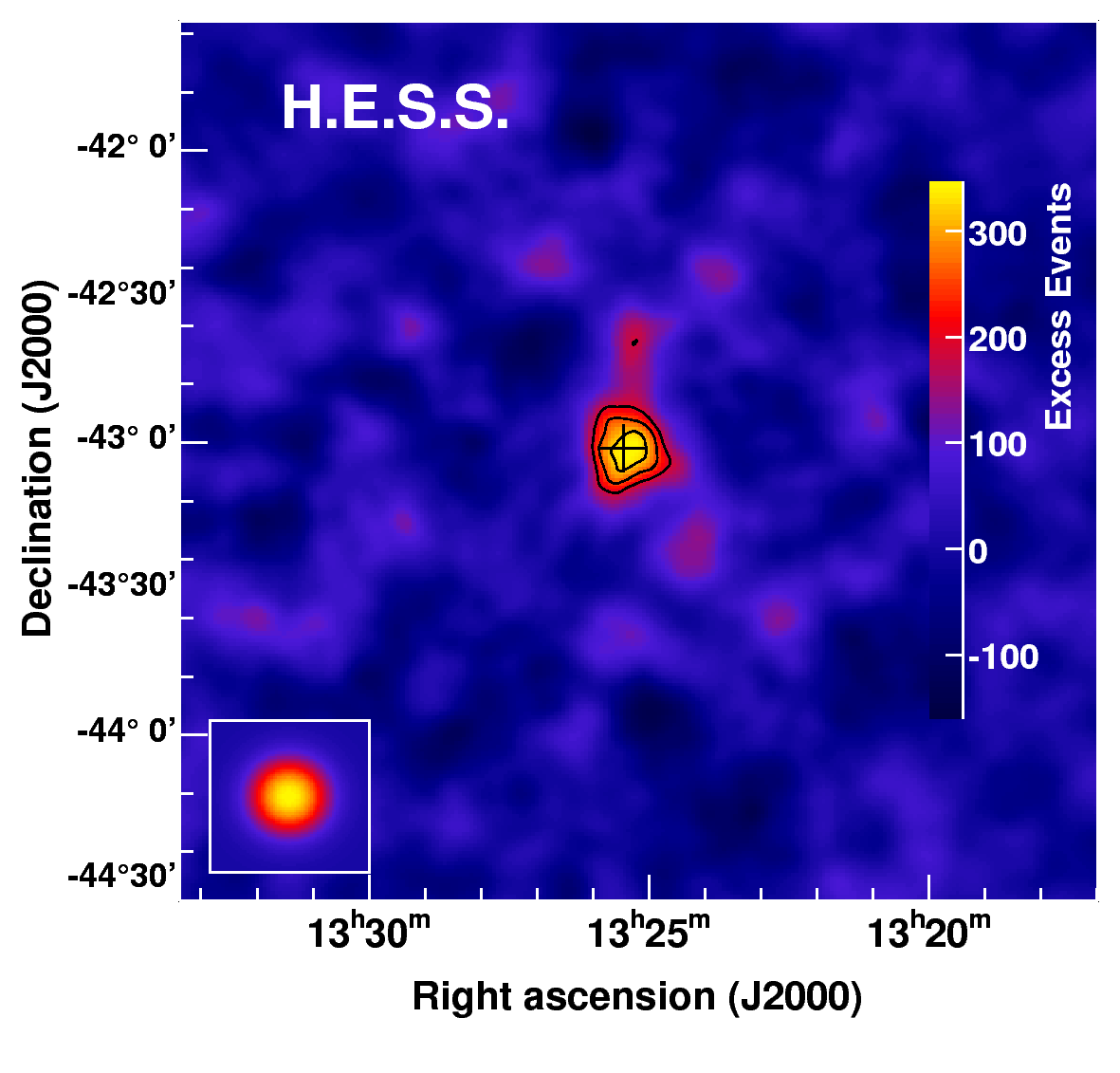}
\caption{Smoothed excess sky map centered on the Cen~A radio core (cross). Overlaid contours correspond to statistical significances of 3, 4, and 5$\sigma$, respectively. The inlay in the lower left corner shows the excess expected from a point source (derived from Monte Carlo simulations). The integration radius is 0.1225\,$^\circ$ and the map has been smoothed with a two-dimensional Gaussian of radius 0.02$^\circ$ to reduce the effect of statistical fluctuations. The cosmic-ray background in each bin is estimated using the template background method \citep{rowell:2003a}.}
\label{Fig:Skymap}
\end{figure}
 
\begin{figure}
\centering
\includegraphics[width=0.36\textwidth]{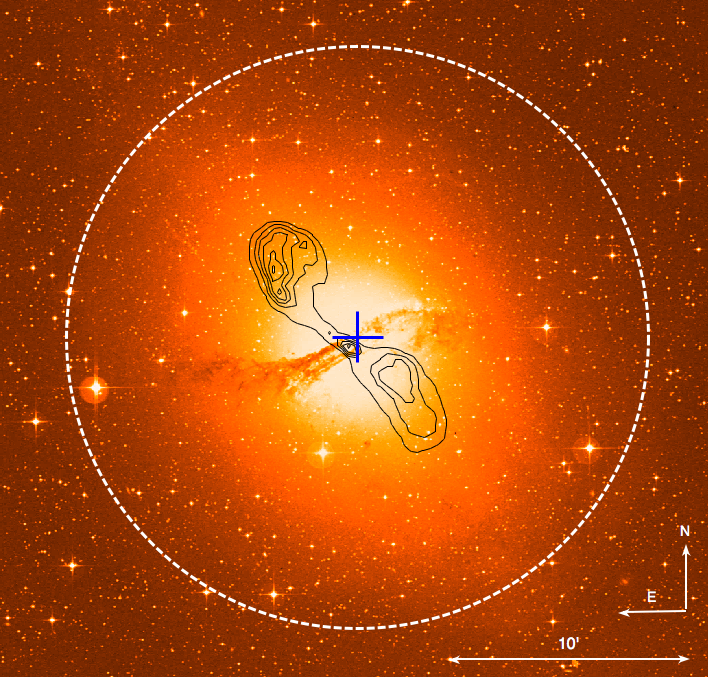}
\caption{Optical image of Cen~A (UK 48-inch Schmidt) overlaid with radio contours (black, VLA, \citealt{condon:1996a}), VHE best fit position with 1\,$\sigma$ statistical errors (blue cross), and VHE extension upper limit (white dashed circle, 95\% confidence level).}
\label{Fig:OpticalRadioVHEMap}
\end{figure}
 
\begin{figure}
\centering
\includegraphics[width=0.5\textwidth]{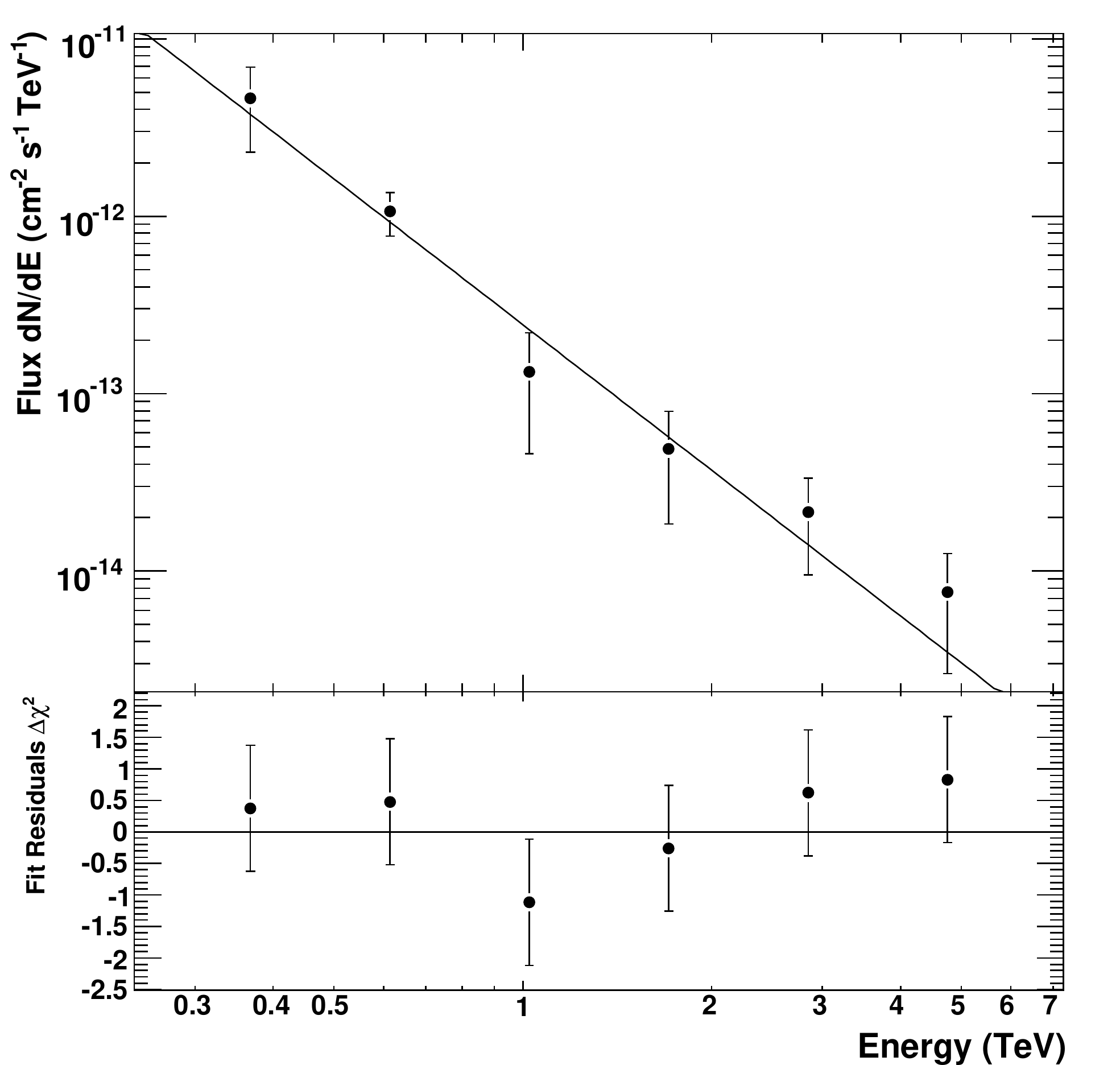}
\caption{Differential energy spectrum of Cen~A as measured by H.E.S.S. The line is the best fit of a power law $dN/dE = \Phi_0 \cdot E^{-\Gamma}$to the data ($\Gamma = 2.7 \pm 0.5_{\mathrm{stat}} \pm 0.2_{\mathrm{sys}}$). The lower panel shows the residual $\Delta \chi^2$ of the fit.}
\label{Fig:Spectrum}
\end{figure}
 
The H.E.S.S. (High Energy Stereoscopic System) collaboration operates an array of four large imaging Cherenkov telescopes (IACT) for the detection of  VHE $\gamma$-rays, located in the Southern Hemisphere in Namibia \citep{aharonian:2006:hess:crab}.
The H.E.S.S. observations of Cen~A were performed between April  2004 and July 2008. A dead time corrected total live time of 115.0\,h of good-quality data was recorded. The zenith angles of the observations range from 20$^\circ$ to 60$^\circ$ with a mean zenith angle of $\sim24^\circ$. The data were recorded with pointing offsets between 0.5$^\circ$ to 0.7$^\circ$ relative to the radio core position, to enable a simultaneous estimation of the background using events from the same field of view (reflected background) \citep{aharonian:2006:hess:crab}. The data were analyzed with a standard Hillas-type analysis \citep{aharonian:2006:hess:crab} with an analysis energy threshold of $\sim250$\,GeV for a zenith angle of 20$^\circ$.

Figure \ref{Fig:Skymap} shows the smoothed excess sky map of VHE $\gamma$-rays as measured with H.E.S.S. centered on the Cen~A radio core position. A clear excess at the position of Cen~A is visible. A point source analysis, using standard cuts as described in \citet{aharonian:2006:hess:crab}, was performed on the radio core position of Cen~A, resulting in the detection of an excess with a statistical significance of $5.0\,\sigma$ (calculated following Eq.~17 of \citealt{li:1983a}; 330 excess events,  $N_{\mathrm{ON}} = 4199$, $N_{\mathrm{OFF}} = 42513$,  $\alpha = 0.091$). A fit of the instrumental point spread function\footnote{Derived from Monte Carlo simulations.} to the uncorrelated sky map results in a good fit (chance probability $\sim0.7$) with a best fit position of  $\alpha_{J2000}=13^{\mathrm h}25^{\mathrm m}26.4^{\mathrm s}\pm4.6^{\mathrm s}_{\rm stat}\pm2.0^{\mathrm s}_{\rm syst}$, $\delta_{J2000}=-43^{\circ}0.7'\pm1.1'_{\rm stat}\pm30''_{\rm syst}$, well compatible with the radio core and the inner kpc jet region (Fig.~\ref{Fig:OpticalRadioVHEMap}). Assuming a Gaussian surface-brightness profile, we derive an upper limit of 0.2$^\circ$ on the extension (95\% confidence level).

The differential photon spectrum of the source is shown in Fig.~\ref{Fig:Spectrum}.\footnote{To derive the energy spectrum, a looser cut on the distance to the source is used ($\theta^2 < 0.03$\,deg$^2$) to increase the number of photons (the standard cut is $\theta^2 < 0.015$\,deg$^2$).} A fit of a power-law function $dN/dE = \Phi_0 \cdot (E/\mbox{1\,TeV})^{-\Gamma}$ to the data is a statistically good description ($\chi^2/\mathrm{d.o.f.} = 2.76 / 4$) with normalization $\Phi_0 = (2.45 \pm 0.52_{\mathrm{stat}} \pm  0.49_{\mathrm{sys}}) \times 10^{-13}$\,cm$^{-2}$\,s$^{-1}$\,TeV$^{-1}$ and photon index $\Gamma = 2.73 \pm 0.45_{\mathrm{stat}} \pm 0.2_{\mathrm{sys}}$. The integral flux above 250\,GeV, taken from the spectral fit, is $\Phi(E > 250\mathrm{\,GeV}) = (1.56 \pm 0.67_{\mathrm{stat}}) \times 10^{-12}$\,cm$^{-2}$\,s$^{-1}$, which corresponds to $\sim 0.8$\% of the flux of the Crab Nebula above the same threshold \citep{aharonian:2006:hess:crab}, or an apparent luminosity of  L($>$250\, GeV)$=2.6 \times 10^{39}$\,erg\,s$^{-1}$ (adopting a distance of 3.8\,Mpc).

No significant variability has been found on time-scales of 28\,min, nights and months (moon periods).
From the error on the nightly flux ($\sim5$\%\,Crab), we estimate the sensitivity of the dataset for flares with a duration of one day. For a $\sim4\,\sigma$ detection in the Cen~A dataset a flux increase during a single night by a factor $\approx20$ would be needed (a factor $\approx15$ for $\sim3\,\sigma$). This can be compared to the VHE flux variation of factor  $\sim5-10$ detected from M\,87 on time-scales of days.

The results have been cross-checked with independent analysis and calibration chains and good agreement was found.

\section{Discussion}
 
\begin{figure}
\centering
\includegraphics[width=0.5\textwidth]{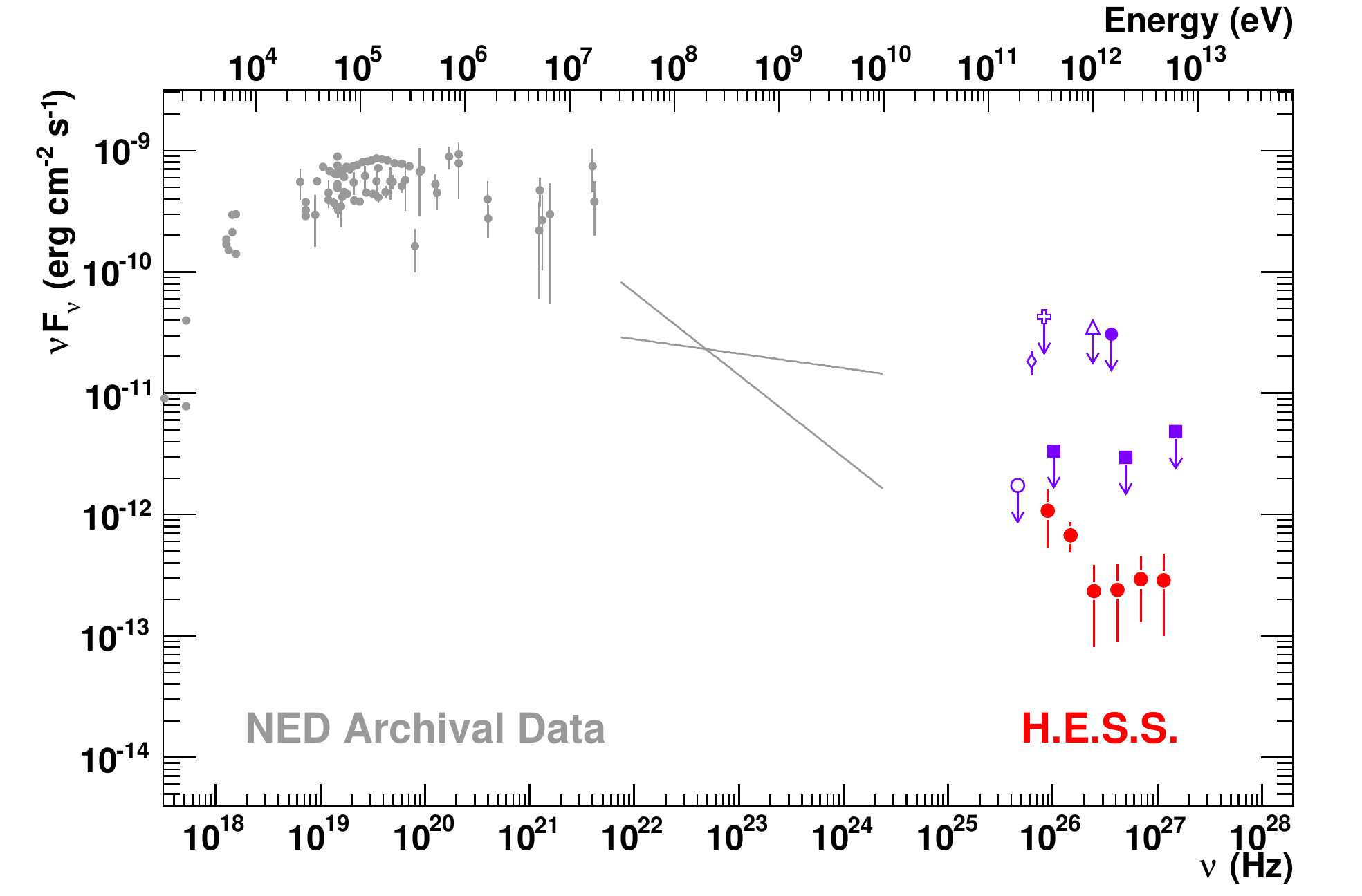}
\caption{Spectral energy distribution of Cen~A. Shown are the VHE spectrum as measured by H.E.S.S. (red filled circles), previous upper limits and tentative detections in the VHE regime (purple markers; \citealt{grindlay:1975a}: open diamond; \citealt{carraminana:1990a}: open cross; \citealt{allen:1993a}: filled circle; \citealt{rowell:1999a}: open triangle; \citealt{aharonian:2005:hess:agnul}: open circle; \citealt{kabuki:2008a}: filled squares), EGRET measurements in the GeV regime (\citealt{sreekumar:1999a}: grey bow tie), and data from the NASA Extragalactic Database (NED) (grey filled circles).}
\label{Fig:SED}
\end{figure}
 
Figure~\ref{Fig:SED} shows the spectral energy distribution of Cen~A ranging from X-rays to the VHE regime.
The flux measured by H.E.S.S. is clearly below all previous upper limits in the VHE regime. Extrapolating the spectrum measured with EGRET in the GeV regime to VHE energies  roughly matches the H.E.S.S. spectrum, though the softer end of the error range on the EGRET spectral index is preferred. The recently launched Fermi observatory should provide a more accurate spectrum in the MeV-GeV range soon.

Several authors have predicted VHE emission from Cen~A, and more generally discussed VHE emission from radio galaxies.
A first class of models proposed the immediate vicinity of the supermassive black hole as the region of VHE emission.
\citet{neronov:2007a} and \citet{rieger:2008a} proposed a pulsar-type particle acceleration in the magnetosphere of the sub-Eddington accreting supermassive black hole, which has been applied successfully to the low-luminosity radio galaxy M\,87. In this model the relation $L_\mathrm{IC} \propto M_\mathrm{BH}$ is expected \citep[see Eq.~(12) in][]{rieger:2008a}. Since the mass of the central black hole and the VHE luminosity from Cen\,A are $\sim 10$ times lower than the ones in M\,87, such a model could possibly be applied to the current data on Cen\,A.

It has also been proposed that, similar to the mechanism at work in other VHE blazars, VHE emission could be produced in the inner jet regions in radio galaxies \citep{bai:2001a,chiaberge:2001a}. Given the jet angle of 15-80$^\circ$ towards the observer \citep[see e.g.][and references therein]{horiuchi:2006a}, the relativistic boosting in standard scenarios would be small. %
\citet{ghisellini:2005a} proposed a two-flow type model \citep[][]{sol:1989a}, with a fast spine and a slower, mildly relativistic sheath propagating within the jet, which has been successfully applied to M\,87 \citep[][]{tavecchio:2008a}. Their model for the SED of Cen~A \citep[see Fig.~3 in][]{ghisellini:2005a}, which was mainly constrained by the available CGRO data, cuts off at lower frequencies than VHE, but could possibly account for H.E.S.S. data by refining the parameters \citep[see also in this context][]{marcowith:1998a}.
\citet{lenain:2008a} modeled the VHE emission of Cen~A with a multi-blob SSC model. In this model, VHE emission is expected to take place in the broadened jet formation zone, where even for a large jet angle, a few emission zones can move directly towards the observer and Doppler boost the emission. Their model prediction for the VHE emission from Cen~A is well compatible with the H.E.S.S. data reported here \citep[see Fig.~7 in ][]{lenain:2008a}.

More extended VHE emission may also be expected from Cen~A.
In this context, \citet{stawarz:2006b} proposed that $\gamma$-rays emitted in the immediate vicinity of the active nucleus are partly absorbed by the starlight radiation in the host galaxy.
The created $e^\pm$ pairs are quickly isotropized and radiate VHE $\gamma$-rays by inverse Compton scattering the starlight radiation. The small size of the resulting isotropic pair halo ($\sim 4$ arcmin in diameter) is fully consistent with a point-like source for H.E.S.S., but could be resolved by the future CTA (Cherenkov Telescope Array)\footnote{http://www.cta-observatory.org/} observatory. Stawarz et al.'s prediction for the VHE emission of Cen~A within this model, resulting in a steady flux of a few $10^{-13}$erg\,cm$^{-2}$\,s$^{-1}$ and a photon index of $\sim 2.6$ at the TeV energy range, agrees well with the H.E.S.S. spectrum \citep[see Fig.~6 in ][]{stawarz:2006b}.

Furthermore, hadronic models have been invoked to predict VHE emission from radio galaxies. In this context, \citet{reimer:2004a} proposed a synchrotron-proton blazar model for M~87 where the high energy component in the SED is interpreted in terms of synchrotron emission from either primary protons or secondary $\mu^\pm$ and $\pi^\pm$ created in the inner jet. At larger distances and in the peculiar case of Cen~A, ultrarelativistic protons with $E_p <10^{18}$\,eV could be released within the kpc-scale jet,
and would be then confined in the host galaxy by interactions with the interstellar magnetic field. Both the dust lane of Cen~A and thermal gas of the merger remnant could act as efficient, dense targets of cold material for these protons, resulting in VHE $\gamma$-ray emission.

\citet{kachelriess:2008a} considered possible UHECR emission from Cen~A in the view of the CGRO observations and current limits provided by imaging atmospheric Cherenkov telescopes.
Depending on the choice for the UHECR injection function in their model, some solutions proposed for the predicted VHE $\gamma$-ray spectrum are compatible with the H.E.S.S. results reported here \citep[see Fig.~1 in ][]{kachelriess:2008a}.

It has also been argued that VHE emission could originate from the outer giant radio lobes \citep[e.g.][]{hardcastle:2008a} lying at $\sim 4^\circ$ from the core. However, the VHE $\gamma$-ray excess presented here only matches the position of the core, the pc/kpc inner jets and the inner radio lobes. Only upper limits at VHE have been derived for the southern outer lobe \citep[][]{kabuki:2008a}.

Recently, \citet{croston:2009a} reported the detection of non-thermal X-ray synchrotron emission from the shock of the southwest inner radio lobe. The edge of this lobe is located $\sim 5'$ from the nucleus and reveals edge-brightened X-ray emission \citep[][]{kraft:2003a}. While the position is $\sim3\sigma$ away from the best fit position of the VHE excess, it is well within the upper limit of the extension. Studying the spatial variation of the spectral index in X-rays across this shock region, \citet{croston:2009a} constrained the high energy cut-off in the electron distribution to be $\gamma_\mathrm{max} \sim 10^8$. They investigated inverse Compton scattering the starlight radiation and the CMB from high energy particles in this lobe and predicted a VHE emission well compatible with the H.E.S.S. measurement reported here (see their Fig.~8) for both seed radiation fields. This study would suggest Cen\,A is analogous to a gigantic supernova remnant (SNR).

Besides the components of the AGN hosted in Cen\,A, one might wonder whether other sources in the host galaxy could be responsible for the VHE emission.
For example, \citet{kraft:2001a} detected more than 200  X-ray point sources in Cen~A, and IACTs have detected about 60 VHE sources in our Galaxy, many of which are associated with SNRs and pulsar wind nebulae (PWNe).
However, source ensembles, such as the sum of the contributions from SNRs/PWNe, would require an unrealistically large number of sources (assuming a typical luminosity of $\sim 10^{34-35}$\,erg\,s$^{-1}$ above 250\,GeV per source).

Further information on the VHE excess position, more detailed spectral shape and variability studies are required to differentiate between the different models. However, if the VHE emission is due to a misaligned blazar-like process -- such as leptonic or hadronic emission from the jet -- the proximity of Cen~A makes it a very good laboratory to further investigate emission processes and jet physics in blazars. If the VHE emission originates from another process -- e.g. UHECR interacting with the interstellar medium, or a SNR type process at the shock -- this would be very exciting, giving new insights into the physics in the VHE domain.

Cen\,A represents a rich potential for future VHE experiments. Our current data are at the edge of differentiating the possible emitting regions. With higher sensitivity (factor 10), better astrometric accuracy and angular resolution (e.g. $\sim 5''$ and $\sim 1'$, respectively) \citep[][]{hermann:2007a}, CTA would allow the localization of the site of the VHE emission, and, possibly, reveal multiple VHE emitting sources within Cen~A. More generally, the detection of VHE emission from Cen~A together with the detection of M\,87 and the galactic center poses the question of whether VHE emission ($\gamma$-ray brightness) might be a general feature of AGN. While the sensitivity of current generation experiments is probably too low to answer this question, one can hope that the CTA experiment will be able to detect a large enough sample of sources to shed some light on this issue.


\acknowledgments
{
\small
The support of the Namibian authorities and of the University of Namibia
in facilitating the construction and operation of H.E.S.S. is gratefully
acknowledged, as is the support by the German Ministry for Education and
Research (BMBF), the Max Planck Society, the French Ministry for Research,
the CNRS-IN2P3 and the Astroparticle Interdisciplinary Programme of the
CNRS, the U.K. Science and Technology Facilities Council (STFC),
the IPNP of the Charles University, the Polish Ministry of Science and 
Higher Education, the South African Department of
Science and Technology and National Research Foundation, and by the
University of Namibia. We appreciate the excellent work of the technical
support staff in Berlin, Durham, Hamburg, Heidelberg, Palaiseau, Paris,
Saclay, and in Namibia in the construction and operation of the
equipment.
This research has made use of the NASA/IPAC Extragalactic Database (NED) which is operated by the Jet Propulsion Laboratory, California Institute of Technology, under contract with the National Aeronautics and Space Administration, and NASA's Astrophysics Data System.
}


\end{document}